\documentclass[aps,twocolumn,showpacs,nofootinbib]{revtex4}
\usepackage{graphicx}

\newcommand{\beq}{\begin{equation}}
\newcommand{\eeq}{\end{equation}}
\newcommand{\beqa}{\begin{eqnarray}}
\newcommand{\eeqa}{\end{eqnarray}}

\def\<{\langle}
\def\>{\rangle}

\newcommand{\complex}{{\kern .1em {\raise .47ex\hbox {$\scriptscriptstyle |$}}\kern -.4em {\rm C}}}
\newcommand{\real}{{{\rm I} \kern -.19em {\rm R}}}

\usepackage{epstopdf}
\begin{document}

\title{Quantum Communication}

\author{Nicolas Gisin and Rob Thew}

\affiliation{Group of Applied Physics, University of Geneva, 1211 Geneva 4, Switzerland}

\date{\today}

\begin{abstract}
Quantum communication, and indeed quantum information in general, has changed the way we think about quantum physics. In 1984 and 1991, the first protocol for quantum cryptography and the first application of quantum non-locality, respectively, attracted a diverse field of researchers in theoretical and experimental physics, mathematics and computer science. Since then we have seen a fundamental shift in how we understand information when it is encoded in quantum systems. We review the current state of research and future directions in this new field of science with special emphasis on quantum key distribution and quantum networks.
\end{abstract}

\maketitle




\section{Introduction}\label{intro}
Quantum communication is the art of transferring a quantum state from one place to another.
Traditionally, the sender is named Alice and the receiver Bob. The basic motivation is that quantum
states code quantum information - called qubits in the case of 2-dimensional Hilbert spaces - and that quantum information allows one to perform tasks that could only be achieved far less efficiently, if at all, using classical information. The best known example is Quantum Key Distribution (QKD)
\cite{BB84,Ekert91,RMP}. Actually, there is another motivation, at least equally
important to most physicists, namely the close connection between quantum communication and quantum non-locality \cite{NonLocalPopescu,BellSpeakable}, as illustrated by the
fascinating process of quantum teleportation \cite{QtelepPRL}.

Quantum communication theory is a broad field, including e.g. communication complexity \cite{Brassard01} and quantum bit-string commitment \cite{Buhrman05}. In this review we restrict ourselves to its most promising application, QKD, both point to point and in futuristic networks. 

There are several ways to realize quantum communication. We list them below from the simplest to the
more involved. Since "flying qubits" are naturally realized by photons, we often write
"photon" for "quantum system", although in principle, any other quantum system could do the job.
\begin{itemize}
\item 1 photon: Alice encodes the state she wants to communicate into a quantum system and sends it to Bob, sections \ref{Qcrypto} \& \ref{securityQKD}.
\item 2 photons: Exploit entanglement to prepare the desired quantum state at a distance, section \ref{entnonLocal}.
\item 3 photons: Teleport the quantum state from Alice to Bob, section \ref{Qtelep}.
\item 4 photons: Teleport entanglement, also called entanglement swapping, section \ref{relays} \& \ref{Qmemory}.
\end{itemize}

We will review quantum communication not with this complexity in mind but from a more intuitive perspective, starting from the basic  ingredient, namely entanglement and its
non-locality, continuing in section \ref{Qcrypto} with weak laser pulse QKD and its security
(section \ref{securityQKD}), before discussing quantum teleportation in section \ref{Qtelep}. We end
by reviewing quantum relays and repeaters (section \ref{relays}), the latter requiring quantum
memories (section \ref{Qmemory}). Along the way, we underline future challenges.

\section{Entanglement \& non-locality}\label{entnonLocal}
Entanglement is the essence of quantum physics. To  understand this statement already  stressed by
Schr\"odinger in 1935 \cite{Schrodinger35a}, it is worth presenting it in  modern terms inspired by quantum information
theory. In Science in general, all experimental evidence takes the form of conditional
probabilities: if observer $A_i$ performs the experiment labelled $x_i$, she observes $a_i$ and in general one writes the probability for  all of the possible results $P(a_1\ldots a_n|x_1\ldots x_n)$. Such conditional probabilities
are often called {\it correlations}.  For simplicity, we restrict the discussion here to the bi-partite case, denoting their correlation $P(a,b|x,y)$.

\begin{figure}
\includegraphics[width=80mm]{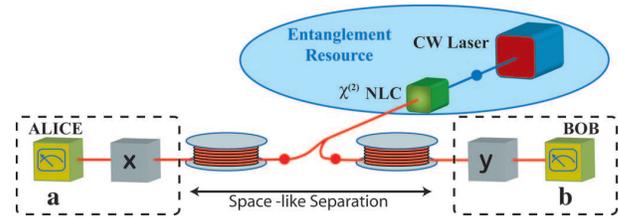}
\caption {\label{fig:nonlocality1} Revealing non-locality. Alice and Bob independently perform experiments x and y, on an entangled state at space like separated locations, and study the correlations for the results a and b.}
\end{figure}

The correlations $P(a,b|x,y)$ carry a lot of  structure. Apart from being non-negative and
normalized, the local marginals are independent of the experiment performed by independent
observers: $\sum_a P(a,b|x,y)=P(b|y)$ is independent of the experiment $x$ performed by Alice. As a
trivial example of independent observers, imagine two physicists performing different experiments in
labs in distant countries, in which case the independence of the marginals is obvious. There is
however another more interesting situation. Suppose the two parties perform similar experiments,
but at two space-like separated locations, thus preventing any communication, as is the case in
Fig.\,\ref{fig:nonlocality1}.  It is therefore natural to assume that the local probabilities
depend only on the {\it local state of affairs} and, as the local state of affairs may be unknown,
one merely denotes them by a generic symbol $\lambda$. Note that the local state of affairs at
Alice's site and at Bob's site may still be correlated. This is why computer scientists call
$\lambda$ {\it shared randomness}. Given the local state of affairs, the correlations factorize to
{\it local correlations}, $P(a,b|x,y,\lambda)=P(a|x,\lambda)\cdot P(b|y,\lambda)$, which
necessarily satisfy some (infinite) set of inequalities, known as Bell Inequalities
\cite{BellSpeakable,Collins04a}. Let us emphasize that there is no need to assume predetermined values to derive Bell Inequalities, it suffices to assume that the probabilities of results of local
experiments depend only on local variables.

Almost all correlations between independent observers known in Science are local. The only
exceptions are some correlations predicted by quantum physics when the two observers perform
measurements on two (or more) entangled systems. This implies that in some cases, a quantum
experiment performed at two distant locations can't be completely described by the {\it local state
of affairs} \cite{BellSpeakable}, a very surprising prediction of quantum physics indeed!

Einstein, among others, was so surprised by this that  he concluded that it "proves" the
incompleteness of quantum mechanics \cite{EPR35}. Following Bohr's reply to the famous EPR paper,
the debate became philosophical. John Bell resolved this with the introduction of the experimental question of Bell Inequalities \cite{BellSpeakable,CHSH,Collins04a} and remarkably, by 1991, it had become applied physics \cite{Ekert91}. Indeed, it was realized that the non-existence of a local state of affairs guaranties that Alice and Bob's data have no duplicate anywhere else in the world, in particular
not in any adversaries' hands. The intuition is clear: since there is no $\lambda$, no one can hold
a copy of $\lambda$, hence no one can compute the probabilities for Alice and Bob's data,
$P(a|x,\lambda)$ and $P(b|y,\lambda)$. Consequently, Alice and Bob's data have some secrecy. This
is the essence of QKD, but clearly, this intuition needs elaboration (see section \ref{securityQKD}). 

Let us conclude this section with a brief review of the experimental and theoretical status of quantum non-locality.
Today, no serious physicist doubts that  Nature exhibits quantum non-locality.  Despite the depth of
such a conclusion (whose revolutionary aspect is often not fully appreciated), it has turned out to
be exceedingly difficult to realize an experiment between space-like separated parties with
detection efficiencies high enough to avoid the detection loophole \cite{GisinGisin99}. While the
detection loophole was closed in an ion trap experiment, the close proximity of the ions ensured
that these were not space-like separated \cite{Rowe01}. Only a couple of experiments have managed
to perform space-like separated tests with entanglement \cite{Aspect82} distributed over ten kilometres both in
fiber \cite{Tittel97,Weihs98,Zbinden01} and free space \cite{Peng05a}, though without closing the detection
loophole. Also on the theory side, it is surprisingly poorly understood why the most well known
Bell inequality, the CHSH-inequality, named after its discoverers \cite{CHSH}, seems the most
efficient one despite the existence of infinitely many other Bell inequalities (however, see \cite{Collins04a}). In particular, we still have no practical way to tell whether a given quantum
state is able to exhibit non-locality or not. This limited understanding is especially
frustrating once one realizes that the experimental violation of a Bell inequality is the {\it only}
direct evidence for the presence of entanglement. Indeed, all the other entanglement witnesses
require that one knows the dimension of the relevant Hilbert space \cite{QcryptoCHSH06}.

\section{Quantum Key Distribution: From Entanglement to Weak Laser Pulses}\label{Qcrypto}
One simple way to think about entanglement for the non specialist is that some composite systems,
like pairs of photons, are able to provide the same random answer when asked the same question. Let
us emphasize that the answer (measurement result) is random, but it is precisely the same
randomness that manifests itself at two distant locations, provided  that Alice and Bob perform the same
experiment (or experiments related by a simple transformation). It then suffices that Alice and Bob
independently choose to perform a series of experiments, drawn from a pre-established list of
possible experiments, and, after recording all their data, they post-select those corresponding
to the cases in which they happened, by chance, to have chosen to perform the same experiment. In
these cases, they asked the same question and thus obtained the same random answer. This provides
them with a cryptographic key. We'll analyze the secrecy of such keys in section
\ref{securityQKD}. In this section we would like to concentrate on practical ways to implement QKD.

The first choice that the quantum telecom  engineer has to face is that of the wavelength. While most quantum optics experiments since the invention of the laser have used silicon-based detectors, limited to wavelengths below 1\,$\mu$m, for long distance quantum communication one should also consider wavelengths suitable for fiber optic communication, 1.3 \& 1.5\,$\mu$m (although space communication to satellites is a serious and fascinating alternative \cite{Aspelmeyer} that we can't review here). Nowadays, there are several options for detectors compatible with optical fibers, ranging from detectors based on superconduction transitions to commercially available APDs (Avalanche PhotoDiodes).

\begin{figure}
\includegraphics[width=80mm]{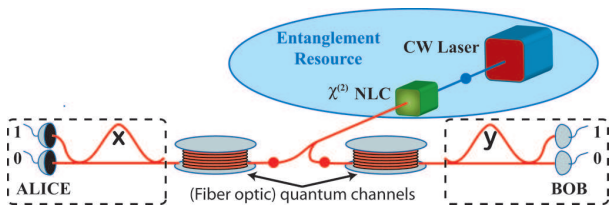}
\caption[efficiency1] { \label{fig:Franson1} The Franson interferometer for testing the energy-time entanglement of the entanglement resource (ER). The correlations between each of Alice and Bob's results \{0,1\} depends on both the phase measurement settings \{x,y\}.}
\end{figure}

The second choice concerns the degree of freedom in which to encode the qubits. An obvious first
choice is the state of polarization, except that polarization is unstable in standard fibers,
especially in aerial fiber cables. In 1989 Jim Franson proposed the use of energy-time entanglement
\cite{Franson89}, with the initial objective to test a Bell inequality, though later adapted to quantum communication. Fig.\,\ref{fig:Franson1} illustrates Franson's idea,  consisting
of a CW laser that pumps a $\chi^2$ nonlinear crystal, where each photon from the pump laser has a
probability of, at best, $~10^{-6}$ to be down-converted into a pair of photons, depending on the
crystal \cite{Tanzilli01}. Each of the two photons has an uncertain energy (i.e. an uncertain
wavelength), where {\it uncertain} should be understood in the quantum mechanical sense. However,
through energy conservation, the sum of the two photon's energy equals the well defined energy of the
pump laser photon. Moreover, both photons are created at the same time (again through energy
conservation), but this time is "quantum uncertain" within the long coherence-time of the pump
laser. We see a nice analogy with the case presented by EPR: the energy and the age of each photon
are uncertain, but the sum of the energies and the difference of their ages are both sharply
defined. Look now at the two unbalanced interferometers and detectors on both sides of
Fig.\,\ref{fig:Franson1}, which have replaced our abstract operations and measurements from
Fig.\,\ref{fig:nonlocality1}, and consider the cases where both photons hit a detector
simultaneously. Recalling that the photons were produced simultaneously, this can happen in two
ways: both photons propagate through the short arm of their interferometers; or both take the long
arms. If the imbalance of both interferometers are alike and much smaller than the pump laser
coherence length, then these two paths are indistinguishable. According to quantum physics, one
should thus add the probability amplitudes and expect interference effects. These are 2-photon
interferences and have been used to violate the Bell CHSH-inequality \cite{brendel92, Kwiat93,
Tittel97}. This configuration is thus suitable for QKD, but it is not practical using today's
technology, hence let's simplify it \cite{BrunnerGisin03}.

\begin{figure}
\includegraphics[width=80mm]{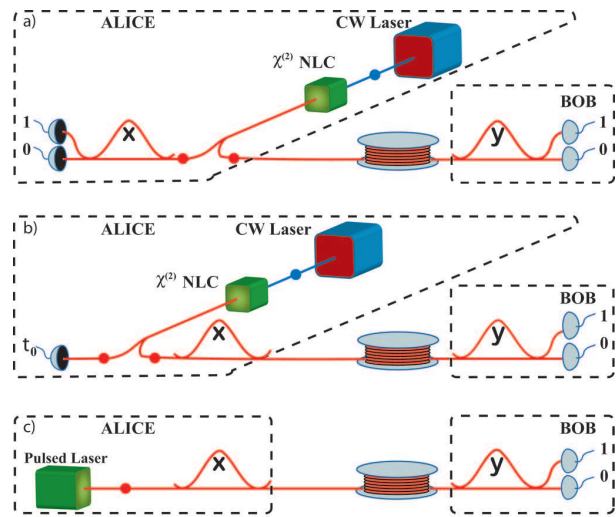}
\caption[efficiency1] { \label{fig:Simplify1}
Simplifying the Franson scheme: a) The ER from Fig.\,\ref{fig:Franson1} is moved to Alice's side; b) The ER is placed before Alice's interferometer - the interfering paths are different but we don't need the extra photon except as a herald; c) Remove ER and replace single heralded photon with attenuated pulsed diode laser.}
\end{figure}

First, let's move the source from the center to the emitter, as in Fig.\,\ref{fig:Simplify1}a, thus
limiting the number of sites to two. Now the photons don't arrive simultaneously at their
detectors but, for an appropriate difference of arrival times, the same reasoning as above applies:
one still has interferences between the short-short and the long-long 2-photon paths. The second
simplification consists of moving the source to the left of Alice's interferometers,
Fig.\,\ref{fig:Simplify1}b. Now the two interfering paths are the short-long and long-short
paths. As before, they are indistinguishable and thus lead to interferences, though now one of the
two photons is not really used (except possibly as a herald). This leads to the third and major
simplification: replace this 2-photon source with a simple weak laser pulse,
Fig.\,\ref{fig:Simplify1}c. The story about the interfering paths remains the same, but the source
is now very simple and reliable: a standard telecom laser-diode with enough attenuation. The 60 to
100 dB attenuation (requiring a well calibrated attenuator) assures that only a very small
fraction of the laser pulses contain more than one photon. It is essential to understand that,
provided this fraction of multi-photon pulses is known, the security of such weak laser pulse QKD
system is in no way compromised \cite{weakPulseSecure,GLLP04}. Moreover, using the recent idea of decoy states, weak laser pulse QKD obeys the same scaling law as ideal single-photon QKD \cite{decoy1,decoy2,decoy3}.

Today, all practical QKD systems use this simplification
\cite{idQ,magiQ,SmartQ,Townsend93,MullerQKDNature95,KarlsonQKD99,Hugues00,BethuneRisk00,
Stucki02,DPS-QKD03,ShieldQKD04,DarpaNetwork05,COW05,Takesue,Thew06} and the
major challenge for QKD (besides the distance, to which we return in section \ref{relays}) is the
secret bit rate. Given that the source is not an issue, there remain two ways to improve this. First,
we can make technical improvements, for example to the detectors whose maximal count rates are
severely limited by dark counts and after-pulses \cite{Ribordy04,Pellegrini}, by using better InGaAs APDs,
up-conversion detection schemes \cite{Thew06,Langrock05} or superconducting detectors
\cite{Goltsman01,Miller03}. Second, the historical protocols, like BB84 and Ekert91
\cite{BB84,Ekert91}, were invented for the sake of presenting a beautifully simple idea, but
today's many new protocols have been designed with the aim of optimizing their implementation using
weak laser pulses \cite{SARGPRL04,decoy1,decoy2,decoy3,decoyHugues,DPS-QKD03,COW05} or mesoscopic systems \cite{Grangier}. It is likely
that more efficient protocols are yet to be discovered by teams combining telecom engineers and
quantum physicists.

\section{Security of QKD}\label{securityQKD} 
The intuition as to why QKD provides perfectly secret bits is quite straightforward (section
\ref{entnonLocal}). However, the details of the proofs are very involved and many
questions remain open, especially concerning optimality \cite{ShorPreskill00,GLLP04,KGR05}.

We would, however, like to highlight just a few key concepts. We can characterize bounds on the
security by comparing  Shannon's mutual information \cite{Shannon} for Alice and Bob $I(A:B)$ and
for Alice and an adversary, traditionally called Eve, $I(A:E)$. It is intuitive (and can be proven
\cite{CsizsarKorner78,Maurer93}) that if Bob has more information than Eve on Alice's data,
$I(A:B)>I(A:E)$, then Alice and Bob can {\it distill} a secret key out of their data. This first
intuition is, however, incomplete.  Eve's information should, in full generality, be treated as
quantum information: there is no way to know whether she performed measurements on her quantum
systems (resulting in classical information) before the key is used. As our goal is to provide a secret key whose security does not rely
on assumptions about Eve's technology, whether classical computer power or quantum technology, this
remark has to be taken seriously. Fortunately, the quantum analog of Shannon's mutual information \cite{RenWol04} and its consequences have recently been resolved \cite{KGR05} .

A second limitation to the above intuitive idea is the so called man-in-the-middle attack: how can
Alice and Bob be sure they really talk to each other? The answer is known and requires that they
start from an initial short common secret, so as to be able to recognize each other. It has been
shown that QKD provides much more secret key than it consumes. In this sense, QKD should be
called {\it Quantum key expansion}.

A third, less studied difficulty are side-channels: how can Alice be sure she doesn't inadvertently
code more than one degree of freedom? For example, it might be that her phase modulator introduces
a measurable distortion of the pulse envelope, in which case Eve could measure the encoded bit
indirectly and remain undetected. A related danger are Trojan horse attacks, in which Eve actively
profits from the quantum channel (i.e. the optical fiber) to probe inside Alice and/or Bob's
systems. Not too much is known to counter such attacks, except by emphasizing that real systems
should be well characterized (see e.g. \cite{Makarov,GisinTrojan06}).

Before we end here, let us briefly elaborate on the widely used terminology {\it unconditionally
secure}. Note that there is nothing like this: security proofs rely on assumptions and some
assumptions are difficult to check in realistic systems. The historical reason for that 
terminology comes from classical cryptography where computer scientists use it to mean "not
conditioned on assumptions about the adversary's classical computation power", a meaning quite
foreign to quantum physics.

\section{Quantum Teleportation} \label{Qtelep} 

\begin{figure}
\includegraphics[width=80mm]{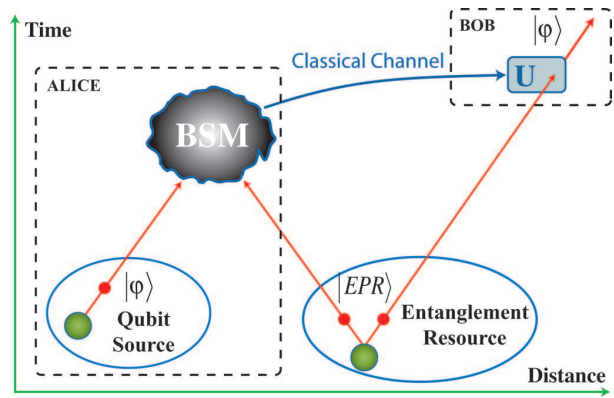}
\caption[efficiency1] { \label{fig:QTeleportation1} Quantum teleportation. Alice performs a BSM, a joint measurement, on the unknown qubit $|\phi\rangle$ and one photon from the entangled state $|EPR\rangle$. The result does not reveal the state of the qubit but is sent to Bob who performs a result-dependent operation U to complete the teleportation.}
\end{figure}

Quantum teleportation is the most fascinating manifestation of quantum non-locality: an "object"
dissolves here and reappears at a distance \cite{QtelepPRL}! Well, not the entire object, "only"
its quantum state, that is its ultimate structure, is transferred from here to there without ever
existing at any intermediate location. The energy-matter must already be present at the receiver
side and must be entangled with the transceiver. Quantum teleportation attracts a lot of attention
from physicists and journalists, and rightly so. Mathematically, quantum teleportation is very
simple, but understanding requires clarifying some often confused concepts
concerning quantum non-locality. 

The entire process requires 3 steps. Consider
Fig.\,\ref{fig:QTeleportation1} where one first has the distribution of entanglement, usually
photon pairs sent through optical fibers (for ions see \cite{QtelepWineland,QtelepBlatt}). The
"quantum teleportation channel" is then established and - in principle - one could remove the
fibers. Next, the sender performs a so-called Bell-State-Measurement (BSM) between his photon from
the entangled pair and the qubit photon that carries the quantum state to be teleported
\cite{BSMWeinfurter94,QtelepBouwmeester97}. Technically, this is the most difficult step and usually
only a partial BSM is realized (see however \cite{QtelepDeMartini98,QtelepShi01}). The BSM provides
no information at all about the teleported state, but tells us something about the relationship
between the two photons \cite{RelativeStateIblisdir06}. 

This ability to acquire information only
about the relationship between two quantum systems is typical of quantum physics: it is another
manifestation of entanglement, but in this case not present between the incoming photons to be
measured. The entanglement lies in the eigenvectors of the operator representing the BSM. Hence,
entanglement plays a dual role in teleportation. Finally, the third step consists of Alice
informing Bob of the result of her BSM and Bob performing a result-dependent unitary rotation on
his system. Only after this operation is the teleportation process finished. Note that the size of
the classical information sent by Alice to Bob is infinitely smaller than the information required
to give a classical description of the teleported quantum state, but it is the need for this
message that ensures that teleportation is a sub-luminal process.

The BSM provides a fundamental limit to these experiments. It has been proven that no BSM with an efficiency greater than 50\%
is achievable with linear optics \cite{BSM50Lutkenhaus}. To perform these partial BSMs, the two
photons should arrive on a beam-splitter simultaneously within their coherence time. Since
single-photon detectors have a large timing jitter, the timing has so far always been set by bulky and
expensive femto-second lasers. Moreover, the length of the optical fibers should be stabilized
within a coherence length of the photons, typically a few tens of microns, an unrealistic
requirement over tens of kilometres. Consequently, some of the next steps will require detectors
with improved jitter \cite{Thew06,Yamamoto06a} as well as compact sources of entangled photons with
significantly increased single-photon coherence. Alternatively, this limitation has been overcome
in some experiments by using continuous variables \cite{QtelepKimble98,QtelepFurusawa98} or
hyperentanglement \cite{QtelepWeinfurter06}, while others have used generalized quantum
measurements to probabilistically distinguish 3 out of the 4 Bell states \cite{BellBrunner06} (it
is an open question whether all 4 could be distinguished using passive linear optics). The intense
interest in BSMs is due to the key role it plays not only in teleportation, but more importantly
its role in long distance quantum communication and specifically entanglement swapping.

 \section{Entanglement Swapping, Relays and Quantum Repeaters} \label{relays}
\begin{figure}
\includegraphics[width=80mm]{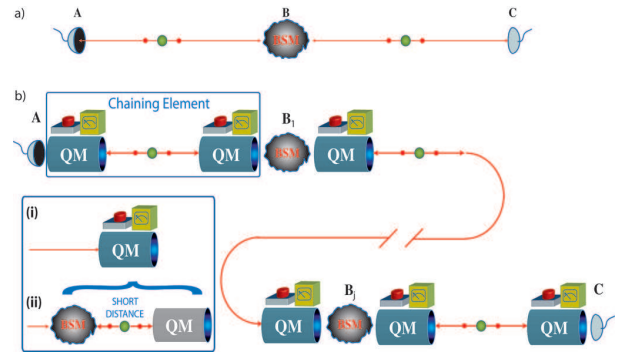}
\caption { \label{fig:QRepeater1} Quantum networks. a) Quantum relay: ERs  and quantum channels joined via a BSM. b) Quantum repeater: An ER + quantum memories (QM) provides a chaining element that can be concatenated for longer quantum communication distances. Inset illustrates the difference between a "heralded QM" (i) and a possible modification for a QM without heralding (ii). }
\end{figure}

What happens if one photon from an entangled pair is teleported, i.e. if entanglement itself is
teleported? This process, known as entanglement swapping, allows one to entangle photons that have
no common past \cite{EntSwapZukowski93}! The general idea consists of first establishing
entanglement between not-too-distant nodes, then teleporting the entanglement from one
node to the next. This is called a {\it quantum relay} \cite{relay1} and the general principle is
illustrated in Fig.\,\ref{fig:QRepeater1}a. So far only very few groups have demonstrated this
process (\cite{EntSwapVienna98,EntSwapVienna02,EntSwapGeneva05}), but this is an active field of
research as it has the potential to increase the distance for QKD.

However, the distances achievable with quantum relays are still limited. The reason is that in
order to be able to swap the entanglement of A-B and B-C to A-C, one first has to establish the
entanglement between A-B and B-C. However, the probability that all photons propagate between A and B and between B and C is precisely the same probability that a photon propagates from A directly to C.
Hence, there is no hope that entanglement swapping by itself helps to increase the bit rate. Still,
quantum relays may be useful for some intermediate distances, because in principle they allow one
to mitigate the detrimental effects of detector dark-counts \cite{relay1,relay2,relay3}.

To efficiently overcome the distance limitation one needs {\it quantum repeaters}, which require
both quantum relays and quantum memories \cite{Briegel98,DLCZ01}. The basic idea is that if the
entanglement distribution has succeeded between nodes A and B, but failed between B and C, one
stores the A-B entanglement in quantum memories and restarts the B-C entanglement distribution. One
can imagine concatenating entangled systems  to further increase this distance (see
Fig.\,\ref{fig:QRepeater1}b). Ideally, one would also like the quantum memories to contain a
rudimentary (few qubit) quantum computer, able to realize the 2-qubit gates for purification or distillation techniques \cite{entConcentration,Deutsch} to concentrate the entanglement contained in each of two pairs of qubits
into a single highly entangled qubit pair. In practice, we are a long way from here but have
started to think about interim possibilities. In a first instance, one may have a  quantum memory where one
doesn't know if it is loaded. In this case  one could place the sources closer to one of the quantum memories in each chaining element of figure \ref{fig:QRepeater1}b.  The motivation behind the asymmetric sources is that if one has one photon directly absorbed by the quantum memory one can be more sure that it is loaded than if it had been transmitted, and possible absorbed/lost in the fiber. This thinking is reminiscent of the simplifications that we made with respect to Fig.\,\ref{fig:Franson1} and the evolution from Franson's intereferometer to weak pulse encoded QKD.

The development of a fully operational quantum repeater and a realistic quantum network architecture are grand challenges for quantum communication. Despite some claims, nothing like this has been demonstrated so far and one should not expect any real-world demonstration for another 5-10 years.

 \section{Quantum Memories} \label{Qmemory} 
If one is to successfully build quantum repeaters then one will need a quantum memory that is able
to store a qubit for a period sufficient to allow several rounds of communication between the
nearby nodes, i.e. typically several ms. In Fig.\,\ref{fig:QRepeater1}b we denote the quantum
memory by some absorbing medium, but more importantly, also with a heralding mechanism so we know when it is loaded. Furthermore, it should either be possible to perform a Bell state measurement between two stored
qubits, or be able to trigger the release of photons carrying the qubits with a jitter small enough
to achieve this, and all of this at wavelengths and bandwidths compatible with existing fiber
optic networks. Today, the best quantum memory by far is a simple fiber loop (though it does not
have all the above mentioned specifications). Storing qubits in some atoms, either in traps or in
some solid-state devices, is a huge challenge. But the potential applications both for fundamental
experiments (e.g. long-distance loophole-free Bell tests) and for a world-wide quantum-web
motivates many physicists. Moreover, it is likely that the successful techniques will also find
applications in other  types of quantum information processors.

Currently there is an increasing number of groups working towards quantum memories from a range of
different perspectives. The different approaches have so far been motivated by the degree of
freedom chosen to encode the quantum state. We have already seen some progress: for continuous
variable systems in atomic vapour \cite{MemoryPolzik}: atomic ensembles \cite{Kimble05,Kuzmich05,Lukin05}; polarization of atom-photon systems
\cite{WeinfurterPhotonAtom}; others are using NV centers in diamonds \cite{Wrachtrup}; as well as
rare-earth ions in fibers and crystals \cite{Kraus06, MansonMemory}. Indeed this last case is
interesting, as most proposals have focused on storing a single mode, or single quantum state,
whereas the rare-earth systems offer the possibility of storing multiple modes, many quantum
states, which could have significant practical implications. These and many more approaches are currently being actively pursued within national and international collaborative programmes around the world \cite{QAPwebPage, SCALAwebPage,EUroadmap,USroadmap}.

\section{conclusion} \label{concl} 
The field of quantum communication has established itself over recent years thanks to its
driving force, Quantum Key Distribution and to the fascinating process of quantum teleportation, not to mention continuous variable \cite{Braunstein05} and satellite quantum communication \cite{Aspelmeyer} and linear optics quantum computation \cite{Myers05}. It will be an important part of physics in the decades to come, with great challenges in quantum memories and repeaters for world-wide applications. It is an ideal teaching tool and is attracting bright young physicists who are learning to build the bridge between quantum physics and communication technologies.

\section*{Acknowledgments}
This work has been supported by the EC under project QAP (contract n. IST-015848) and by the
Swiss NCCR {\it Quantum Photonics}.

\end{document}